\documentclass[reprint,amsmath,amssymb,prb]{revtex4-1}
\usepackage{braket}
\usepackage{bm}

\pdfoutput=1
\usepackage[pdftex]{graphicx}

\begin{document}
\preprint{}


\title{Accurate estimation of interface phase diagram from STM images}

\author{Kazuhito Takeuchi}
\author{Koretaka Yuge}
\author{Shinya Tabata}
\author{Hiroki Saito}
\author{Shu Kurokawa}
\author{Akira Sakai}
\affiliation{Department of Materials Science and Engineering, Kyoto University, Sakyo, Kyoto 606-8501, Japan}


\date{\today}

\begin{abstract}

We propose a new approach to constructing a phase diagram using the effective Hamiltonian derived only from a single real-space image produced by scanning tunneling microscopy (STM).
Currently, there have been two main methods to construct phase diagrams in material science: {\it ab initio} calculations and CALPHAD with thermodynamic information obtained by experiments and/or theoretical calculations.
Although the two methods have successfully revealed a number of unsettled phase diagrams, their results sometimes contradicted when it is difficult to construct an appropriate Hamiltonian that captures the characteristics of materials, e.g., for a system consisting of multiple-scale objects whose interactions are essential to the system's characteristics.
Meanwhile, the advantage of our approach over existing methods is that it can directly and uniquely determine the effective Hamiltonian without any thermodynamic information.
The validity of our approach is demonstrated through an Mg--Zn--Y long-period stacking-ordered structure, which is a challenging system for existing methods, leading to contradictory results.
Our result successfully reproduces the ordering tendency seen in STM images that previous theoretical study failed to reproduce and clarifies its previously unknown phase diagram.
Thus, our approach can be used to clear up contradictions shown by existing methods.

\end{abstract}

\maketitle


Exploring new high-functional materials has remained a greatly challenging theme for several decades.
Phase diagrams play very important roles in their manufacture.
To obtain phase diagrams, both experiments and simulations have to be used.
One can roughly classify the methodologies based on simulations into two main types: one is CALPHAD\cite{10.1038/srep02731, PhysRevB.71.094206} using thermodynamic information obtained by experiments and/or other simulation methods, and the other is {\it ab initio} calculations without experiments.
In particular, for estimating alloy phase diagrams based on microscopic information, {\it ab initio} calculations based on density functional theory (DFT)\cite{PhysRevB.59.1758, PhysRevB.54.11169, PhysRevB.47.558, VANDEWALLE2002539, vandeWalle2002} are now most widely used to determinie the interactions of a given system.
DFT has been applied to bulk, surface, and interface systems and has achieved a number of successful results\cite{PhysRevB.88.094108, 0034-4885-71-4-046501, PhysRevB.71.054102}.

Meanwhile, there remain challenging systems where the results of {\it ab initio} calculations contradict with those of experiments.
Such contradictions can particularly occur in systems where long-range and/or many-body interactions between multiple-scale objects (e.g., atoms and coarse-grained particles) are important because it is typically difficult to determine an effective multiscale Hamiltonian that appropriately describes the behavior of structural phase transitions.
For DFT, it can be practically unfeasible to obtain an accurate Hamiltonian, especially when long-range interactions on a coarse-grained scale come into play, mainly due to the restricted size of the cells used;
for experiments, essential difficulties arise in dividing thermodynamic information into individual mutiscale interactions.
One representative system, that includes such contradictions is Mg-based long-period stacking-ordered structure (LPSO), which has recently attracted a great deal of attention due to its considerable mechanical performance\cite{ZHU20102936, XU201293, 10.1038/srep40846}.
At the stacking-fault in LPSO, so-called nano-sized ``clusters'' consisting of multiple elements are arranged in Mg solvent, and this interface plays a central role in stabilizing the whole LPSO.
Through {\it ab initio} calculations and Monte Carlo simulation, Kimiduka {\it et al.}\cite{LPSO_Kimizuka} determined the multiscale Hamiltonian with pair-wise intercluster interaction energies and revealed that the in-plane nanocluster-ordering tendency is mainly due to the repulsive interaction between nanoclusters.
Although they partially explained the radial distribution function of clusters, the microscopic ordering tendency seen in STM images was not quantitatively reproduced\cite{S.Kurokawa:private}.
Minamoto {\it et al.}\cite{Satoshi_Minamoto2015MH201418} also constructed ternary bulk Mg--Zn--Y phase diagram using CALPHAD.
However, since it is difficult to measure the thermodynamic information of a multiscale system, experiments to determine its Hamiltonian are hard to perform.
Moreover, CALPHAD cannot not show geometrical information concerning the ordering of nanoclusters.

In this study, we present a new approach to constructing a phase diagram by combining our recently proposed theory\cite{2017arXiv170407725Y} with two-dimensional STM images.
Since our approach directly determines underlying many-body interactions only from the geometrical information of any-sized objects at thermodynamic equilibrium within a given accuracy, we can clarify that underlying interactions play an essential role in stabilizing the measured microscopic structure.
Thus, when contradictions occur between {\it ab initio} calculations and experiments, it is expected that our approach, which does not require any thermodynamic information, can bridge this gap.
To demonstrate the validity of our approach, the stacking-fault interface of Mg--Zn--Y alloy, in which ${\rm L1_2}$-type ${\rm Zn}_6{\rm Y}_8$ clusters are arranged, is a suitable target.
We herein show the interface phase diagram of the 18R-type LPSO phases in Mg--Zn--Y alloy, compare our results with previous studies, and successfully reproduce the short-range-order (SRO) of clusters seen in the STM image, which the previous work failed to reproduce quantitatively.

\begin{figure}
	\centering
	\includegraphics[width=0.8\columnwidth]{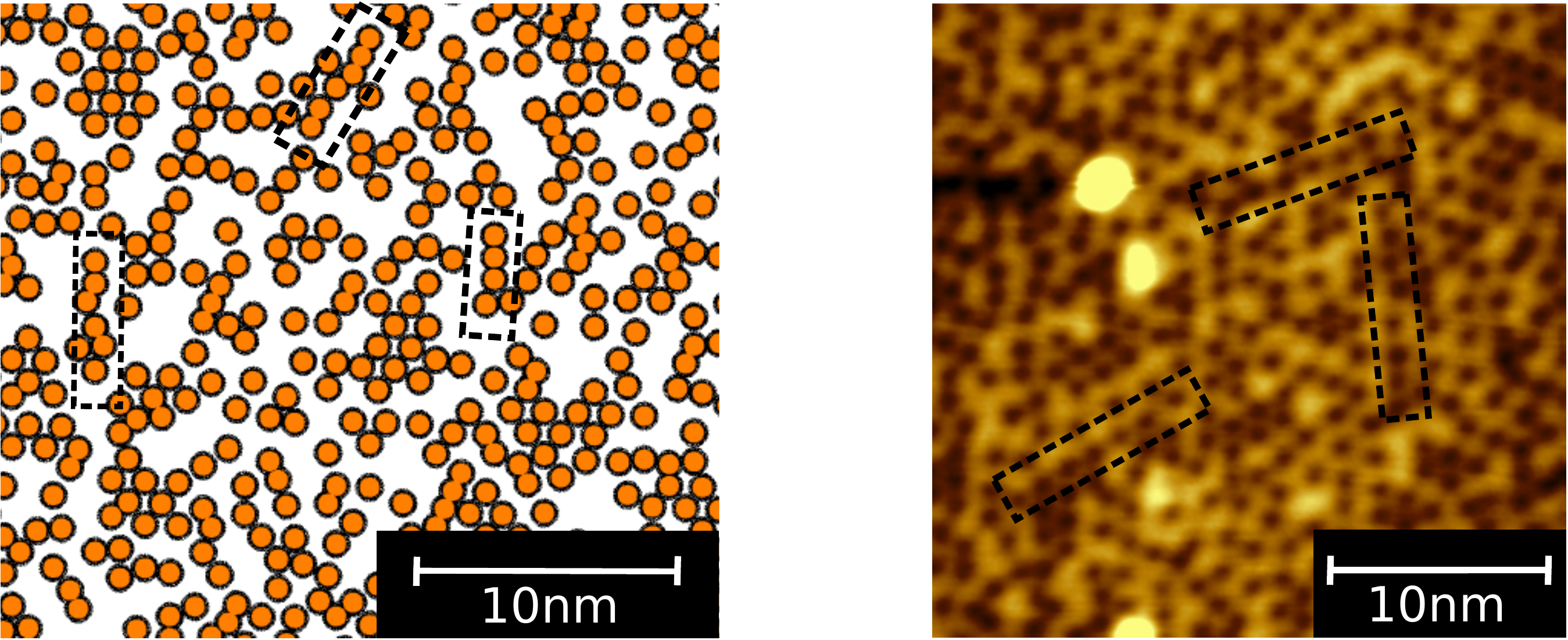}%
	\caption{\label{fig:snapshot}
		(Left) A snapshot of a Monte Carlo simulation at $x = 0.0511$ and $T=773$ K. For clarity, only the Zn--Y clusters are shown.
		(Right) A typical STM image of the distribution of Zn--Y clusters in the stacking fault interface. Individual Zn--Y clusters are imaged as dark spots under positive bias voltage (empty-state image).
		In both images, chain-like arrangements of clusters are denoted by broken-line squares.
	}
\end{figure}
On the right-hand side of Fig.~\ref{fig:snapshot}, we show a typical STM image of the distribution of Zn--Y clusters in the stacking-fault interface (see Methods for sample preparation).
Since the STM is only sensitive to the topmost surface atoms, one can directly determine the individual positions of the clusters\cite{Shu_Kurokawa2013M2013123}.
Before the evaluation of the SRO, we corrected the distortion of STM images, which originates from thermal drift.
In particular, we calculated the vectors between all clusters and selected those corresponding to the 6th-nearest-neighbor (6NN) distance.
STM images were corrected by linear transformation such that 6NN vectors had directions and magnitudes that were as correct as possible. Once we eliminated the effects of thermal drift, we could measure the relative positions between any pairs of clusters. We counted the number of cluster--cluster, cluster--vacancy, and vacancy--vacancy pairs in the field of the STM image up to a distance of 25NN and evaluated SRO.

\begin{figure}
	\centering
	\includegraphics[width=0.8\columnwidth]{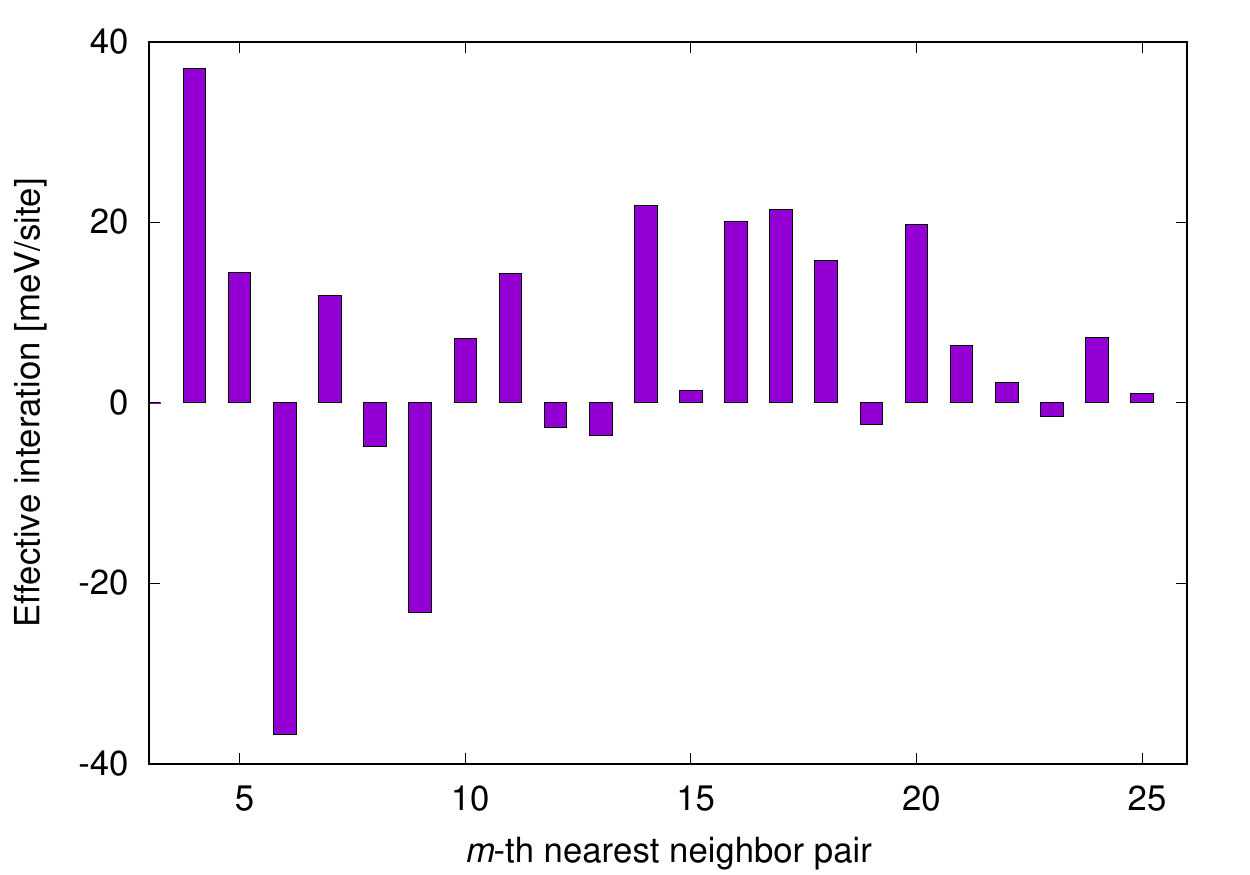}%
	\caption{\label{fig:eci}
		Effective interactions estimated by geometrical information, i.e., the SROs of Zn--Y clusters evaluated from a single STM image on the right-hand side of Fig.~\ref{fig:snapshot}.
	}
\end{figure}
Effective interactions of $m$NN pairs, $V_m$, as determined by SRO from STM images (see Methods) are shown in Fig.~\ref{fig:eci}.
Since in 18R-type LPSO, the interplane interactions are relatively weak\cite{KISHIDA2015228}, this method is valid for reproducing the characteristics of an STM image, ignoring interplane interactions and regarding the interface as an isolated two-dimensional system, i.e., all interactions in Fig.~\ref{fig:eci} are in-plane interactions.
Note that, from our definition of $V_m$, negative $V_m$ implies that $m$NN Mg--Mg and cluster--cluster pairs are favorable and $m$NN Mg--cluster pairs are unfavorable, whereas positive $V_m$ implies opposite.
For example, $V_{4{\rm NN}}$ and $V_{5{\rm NN}}$ show a positive value, meaning that 4NN and 5NN cluster--cluster pairs are unfavorable.
Meanwhile, $V_{6{\rm NN}}$ shows a strong negative value, meaning that 6NN cluster--cluster pairs are favorable, and therefore, an ordering tendency is expected at the interface such that clusters are arranged into each other's 6NN position.

However, our result seems to contradict the previous DFT result\cite{LPSO_Kimizuka}, which held that 6NN cluster--cluster pairs are unfavorable.
As stated above, we actually expect that the most probable reason for this disagreement would originate from differences in our definition of interactions and insufficient consideration of the dependence on the size of a simulation cell in the previous work.
In order to confirm our assumption, using our interactions, we investigated the pair-formation energy of a single $m$NN ($m=6,7,8$) cluster--cluster pair with an $L \times L$ two-dimensional triangular lattice under a periodic boundary condition.
The pair-formation energy of $\Delta E(m,L)$ is defined as
\begin{equation}
	\label{eq:formation_E}
	\Delta E(m,L) = E(m,L) + E_{\rm Mg} - 2 E_{\rm cluster}(L).
\end{equation}
Here, $E_{\rm Mg}$ denotes the total energy of a configuration filled with Mg, and $ E_{\rm cluster}$ denotes total energy of a configuration consisting of a single cluster and  Mg atoms on the rest lattice points in a considered cell.
The definition of Eq.~(\ref{eq:formation_E}) corresponds with that used in the previous work and reflects actual $m$NN pair interactions.
\begin{figure}
	\centering
	\includegraphics[width=0.8\columnwidth]{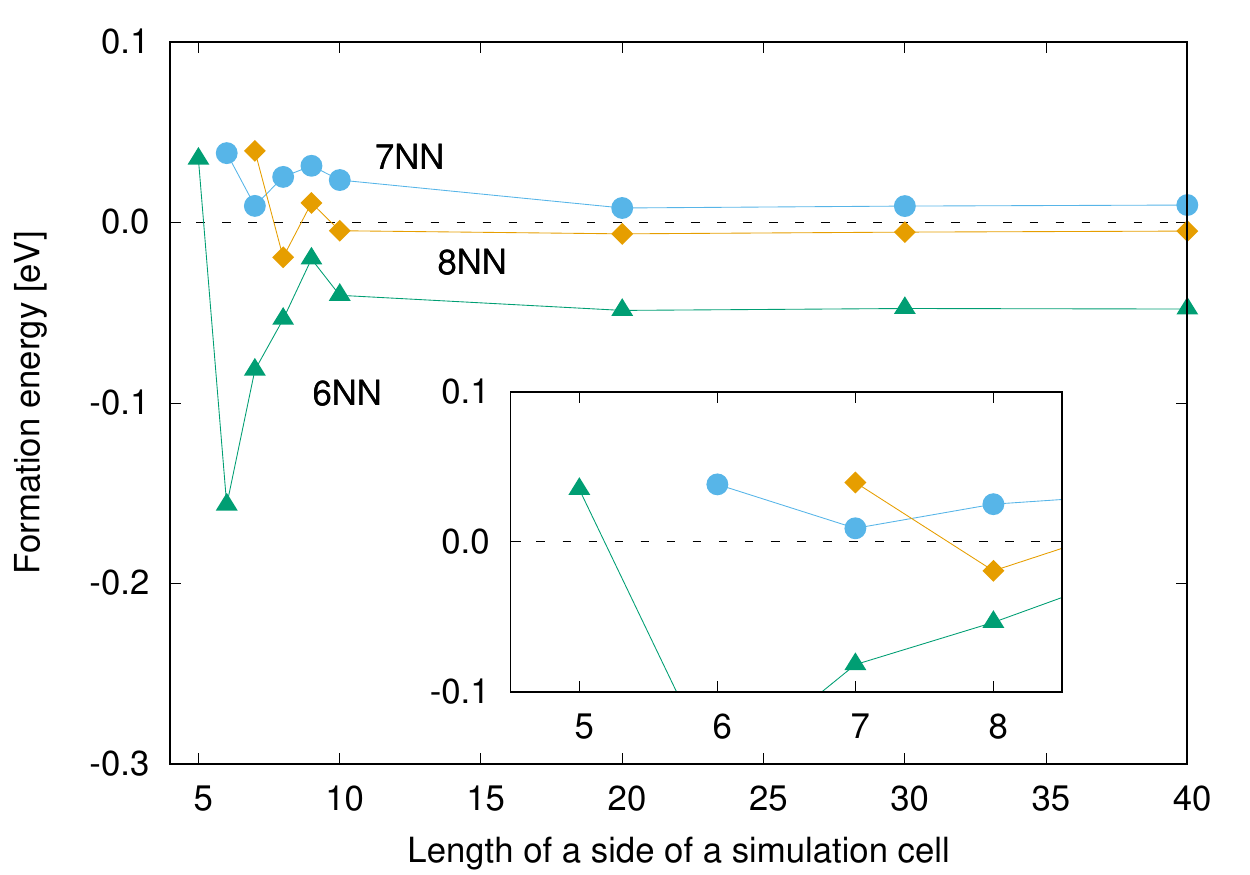}%
	\caption{\label{fig:pair_e_vs_l}
		Pair-formation energy of a single $m$NN pair, as defined by Eq.~(\ref{eq:formation_E}), for each $L \times L$ two-dimensional triangular lattice.
		The inset emphasizes the behavior of pair-formation energy in $L=\hbox{5--8}$ simulation cells.
	}
\end{figure}
In Fig.~\ref{fig:pair_e_vs_l}, $\Delta E(m,L)$ is shown for $m=\hbox{6--8}$ and $L=\hbox{5--40}$.
We found strong oscillation of $\Delta E(m,L)$ in $L \leq 10$, meaning that interactions depend on the size of the cell used, i.e., composition.
In particular, positive values of $\Delta E(6,5)$, $\Delta E(7,6)$ and $\Delta E(8,7)$ imply that the composition of clusters is very dense that cluster--cluster interaction behaves repulsively.
This tendency would be similar to the result obtained by a previous DFT work, which estimated interactions using a simulation cell with fully arranged clusters.
However, the method we used to determine the interactions is different from that used by the previous DFT work.
Meanwhile in a large simulation cell where $L \geq 10$, $\Delta E(6,L)$ becomes strongly negative, unlike $\Delta E(7,L)$ or $\Delta E(8,L)$.
Therefore, the 6NN cluster--cluster interaction behaves attractively for dilute clusters.

\begin{figure}
	\centering
	\includegraphics[width=0.8\columnwidth]{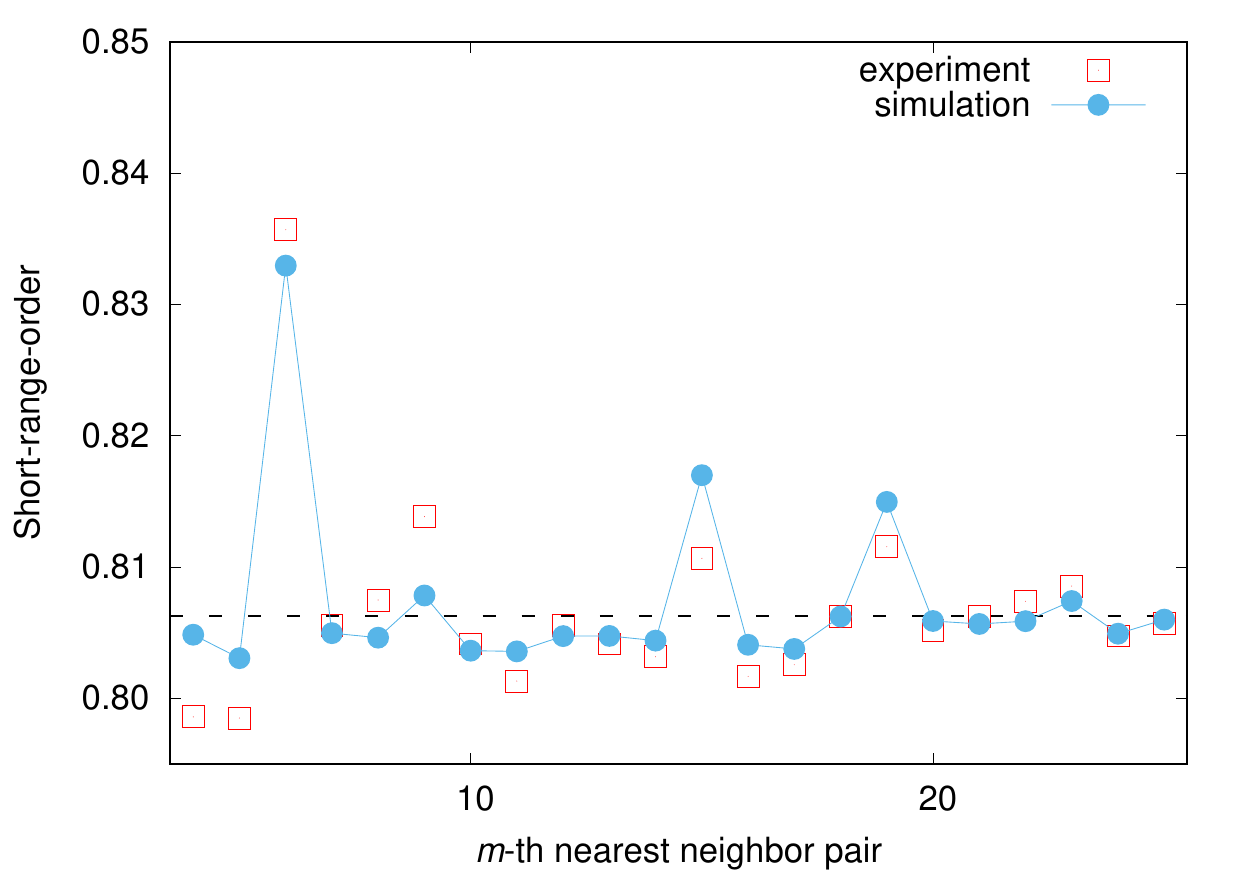}%
	\caption{\label{fig:sro}
		The SRO parameter of ${\rm Mg}_{0.9489}({\rm Zn}_6 {\rm Y}_8)_{0.0511}$ for each $m$th-nearest-neighbor pair at $T=773$ K.
		The broken line denotes the statistical average.
		If SRO is higher than average, the number of Mg--Mg and cluster--cluster pairs is larger than the number of Mg--cluster pairs.
		The opposite is true if SRO is lower than average.
	}
\end{figure}

Although the previous result\cite{LPSO_Kimizuka} successfully showed qualitative landscapes of the radial distribution function of nanoclusters, it could not reproduce the quantitative, microscopic ordering tendency seen in STM images (solid lines in Fig.~\ref{fig:snapshot}).
This is because the previous DFT work determined cluster--cluster interactions by dense-cluster simulation cells, whereas we found that  interactions behave differently between dense and dilute cluster concentrations, as shown in Fig.~\ref{fig:pair_e_vs_l}.
Now, let us confirm that our multiscale interactions in Fig.~\ref{fig:eci} can capture the characteristics of SRO of clusters at the interface.
Fig.~\ref{fig:sro} shows thermodynamically averaged SRO at equilibrium , as obtained by Monte Carlo simulation (see Methods for a more detailed procedure), and this SRO corresponds to the radial distribution function of clusters.
The strong SRO of 6NN pair means that the number of Mg--Mg and cluster--cluster pairs are relatively larger than that of Mg--cluster pairs, i.e., clusters are arranged so as to be at each other's 6NN position.
Our SRO landscape agrees well with the radial distribution functions obtained by previous DFT and experimental work, especially for the peaks in 6, 9, and 15NN pairs.
For further discussion, we took a snapshot at the interface in the simulation and compared it with an STM image in Fig.~\ref{fig:snapshot}.
From the STM image, we can clearly confirm chain-like cluster ordering, unlike the previous DFT work, which showed a uniform arrangement of  clusters.
As above, our snapshot quantitatively captures the features of STM images and supports the validity of our multiscale interactions.

\begin{figure}
	\centering
	\includegraphics[width=0.8\columnwidth]{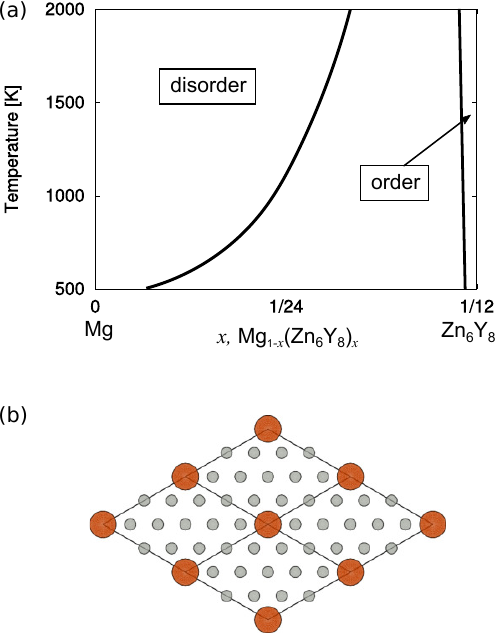}%
	\caption{\label{fig:pd1}
	(a) ${\rm Mg}_{1-x}({\rm Zn}_6 {\rm Y}_8)_x$ interface phase diagram. The broken line denotes the phase-coexistent line between the ordered and disordered phases.
	(b) Ordered phase of the interface. Large and small circles denote Zn--Y clusters and Mg, respectively.
	}
\end{figure}

The ${\rm Mg}_{1-x}({\rm Zn}_6 {\rm Y}_8)_x$ interface phase diagram is presented in Fig.~\ref{fig:pd1}(a), which shows the ordering tendency in the cluster-rich phase (Fig.~\ref{fig:pd1}(b)) through a first-order order-disorder phase transition.
Since, the interplane interactions are relatively weak in 18R-type LPSO\cite{KISHIDA2015228}, it is valid to regard the interface of LPSO as an isolated two-dimensional system.
Compared with the Mg--Y--Zn ternary bulk phase diagram\cite{S.Miura:private} obtained via experiment, the temperature-composition region of 18R-type LPSO shows good agreement between our approach and experiment.

In summary, we suggested a new approach based on microscopic geometric information for constructing phase diagrams in cases where this is difficult for conventional simulation methods.
Using STM images, we demonstrated our approach through the interface of 18R-type Mg--Y--Zn LPSO.
We clarified the contradictory information concerning cluster arrangements presented by experiments and simulations and successfully reproduced an interface phase diagram consistent with that obtained by experiment.
Our approach is expected to become a powerful tool for modeling isolated systems subjected to experimental observations such as STM.

\section{Acknowledgments}
This work was supported by a Grant-in-Aid for Scientific Research (16K06704), Grants-in-Aid for Scientific Research on Innovative Area Nos. 24109503 26109711 from the MEXT of Japan, Research Grant from Hitachi Metals $\cdot$ Materials Science Foundation, and Advanced Low Carbon Technology Research and Development Program of the Japan Science and Technology Agency (JST).

\section{Author contributions}
S. Kurokawa and A. Sakai conceived and designed STM experiments and estimated short-range-order of clusters.
H. Saito prepared the samples and took STM images.
S. Tabata did data processing of STM images and determined the position of the clusters.
K. Yuge modeled the effective Hamiltonian via the method with geometrical information.
K. Takeuchi calculated thermodynamic properties and constructed the phase diagram using the effective Hamiltonian.

\section{methods}

\subsection{Sample preparation for STM images}
The directional solidification process was applied to a ${\rm Mg}_{85}{\rm Zn}_6{\rm Y}_9$ ingot. The ingot was annealed at 773 K for 168 h and quenched with water. We confirmed the presence of a diffraction spot corresponding to 18H-type LPSO by TEM observation.

Samples for STM observation were prepared by cutting the alloy ingot, typically into a reed shape of 8mm $\times$ 5mm $\times$ 0.5mm. After the introduction of the sample into the preparation chamber of the low-temperature ultrahigh-vacuum STM (Unisoku 1200), we cooled the sample with a liquid-nitrogen flow for 10--15 min.  Finally, we cleaved the sample using a pushing rod in the UHV chamber. The cleaved sample was immediately transferred into a pre-cooled observation chamber. All STM observations were carried out at liquid-nitrogen temperature (up to 77 K).

\subsection{Generalized Ising model}
In recent alloy studies with {\it ab initio} calculations, the generalized Ising model\cite{SDG} (GIM) has often been used for constructing a coarse-grained Hamiltonian, which captures the characteristics of a system within a given accuracy.
Since we are interested in how clusters would be arranged at the interface with Mg solvent, for simplicity, it is natural to consider the clusters as coarse-grained atoms, i.e., Mg and the center of cluster as +1 and -1 Ising-spin variables, respectively.
In GIM, the total energy of a given configuration, $\sigma$, is expanded by a set of complete and orthogonal basis functions, $\{ \xi_k \}$, with indices
\begin{equation}
	E^{(\sigma)} = \sum_k V_k \xi_k^{(\sigma)},
\end{equation}
where $V_k$ denotes an effective interaction and $k$ is an index specified as a symmetrically nonequivalent figure such as 1NN and 2NN pairs.
Note that in a binary system, $\xi_k$ corresponds to the average product of spin variables $s_\alpha$ on $k$ over all symmetrically equivalent $k$ in an Ising-spin configuration
\begin{equation}
	\label{eq:xi}
	\xi_k^{(\sigma)} = \sum_{k \in \sigma} \prod_{\alpha \in k} s_\alpha,
\end{equation}
where $\alpha$ denotes a site on a given lattice.

\subsection{Short-range-order in our simulation}
We calculated the SROs of clusters using the Metropolis algorithm with a $100 \times 100$ two-dimensional triangular lattice under fixed composition.
Note that, to explicitly consider the size of the nano-sized clusters without loss of validity, interactions of 1--3NN pairs are configured as having a very high so as not to overlap with each other.

\subsection{Determination of many-body interactions from a single STM image}
We first assume that the measured structure obtained by STM is in thermodynamic equilibrium.
Then, the expectation value of the $k$th coordination of the structure at temperature $T$, $\Braket{\xi_k}\left( T \right)$,
can be given in the framework of classical statistical mechanics as
\begin{eqnarray}
\Braket{\xi_k}\left( T \right) = Z^{-1} \sum_\sigma \xi_k^{\left( \sigma \right)} \exp\left( -\frac{E^{\left( \sigma \right)}}{k_{\textrm{B}}T} \right),
\end{eqnarray}
where summation is taken over all possible configurations $\sigma$, and $Z$ denotes a partition function.

Under these conditions, we have recently derived the relationship\cite{2017arXiv170407725Y} between the structure, $\mathbf{Q}\left( T \right)=\left\{\Braket{\xi_1}\left( T \right),\cdots, \Braket{\xi_f}\left( T \right)\right\}$ and the many-body interaction, $\mathbf{V}=\left\{ V_1,\cdots, V_f \right\}$, in an explicit matrix form:
\begin{eqnarray}
\label{eq:emrs}
&&\mathbf{Q}_{\textrm{ave}} + \bm{\Gamma}\left(T\right)\cdot \mathbf{V} \simeq \mathbf{Q}\left(T\right)  \nonumber \\
&&\Gamma_{ik}\left(T\right) = -\frac{1}{k_{\textrm{B}}T} S_{ik},
\end{eqnarray}
where $\mathbf{Q}_{\textrm{ave}}$ represents a linearly averaged configuration and $S_{ik}$ denotes an element of the covariance matrix for the configurational density of states in $\left( \xi_i,\xi_k \right)$ two-dimensional space for a non-interacting system.
Both of these quantities can therefore be known \textit{a priori} without any information about energy or temperature.
From Eq.~(\ref{eq:emrs}), we can thus directly determine many-body interactions from a measured structure using the equation
\begin{eqnarray}
\mathbf{V} \simeq \bm{\Gamma}^{-1} \cdot \left( \mathbf{Q}\left(T\right) - \mathbf{Q}_{\textrm{ave}} \right).
\end{eqnarray}

\subsection{Constructing the interface phase diagram}

Recently, we proposed a new method based on the Wang-Landau algorithm to construct an alloy phase diagram.
Since the Wang-Landau algorithm can accurately estimate phases that conventional Monte Carlo methods overlook, this method is suitable for estimating the interface phase diagram.
In this study, the interface is regarded as Mg--cluster pseudo-binary alloy.
To estimate its phase stability, we only focus on the chemical potential, $\mu$, between chemical potentials for each constituent element, $\mu_{\rm Mg}$ and $\mu_{\rm cluster}$, $\mu = \mu_{\rm cluster} - \mu_{\rm Mg} $.

To construct the alloy phase diagram, a semi-grand-canonical (SGC) ensemble is more often used than grand-canonical or canonical ensemble.
In an SGC ensemble where compositions can vary when the number of atoms, $N$, kept fixed, the total energy, $E_{\rm SGC}$, is defined as $ E_{\rm SGC} = E - \mu N x$. The free energy, $\phi$, is defined as $\phi = - k_{\rm B} T \ln{Y} $, with Boltzmann constant $k_{\rm B}$.
The partition function $Y$ is defined as
\begin{equation}
	\label{eq:sgc}
	Y (T, \mu)
	= \sum_{E_{\rm SGC}} W (E_{\rm SGC}) \exp \left( -\frac{E_{\rm SGC}}{k_{\rm B}T} \right)
\end{equation}
where $W$ denotes density of states (DOS).
Note that Helmholtz free energy $F$ is related to $\phi$ through Legendre transformation:
\begin{equation}
	\label{eq:legendre}
	\phi = F - \mu Nx.
\end{equation}
We can obtain $x$ using partial differentiation by interpolating $\phi$ for each chemical potential:
\begin{equation}
	\label{eq:composition_sgc}
	x = - \frac{\partial \phi}{\partial (\mu N)}.
\end{equation}
The above procedure compensates for a disadvantage of the Wang-Landau algorithm, which cannot not directly calculate a thermodinamically averaged value (unlike the Metropolis algorithm).
We estimated DOS for a $30 \times 30$ two-dimensional triangular lattice with single-spin-flip update so as to prevent nanoclusters from overlapping with each other.

\bibliography{lpso.bib}

\end{document}